\documentclass[onecolumn,preprintnumbers,superscriptaddress]{revtex4}
\usepackage{amsfonts}
\usepackage{amsmath}
\usepackage{amssymb}
\usepackage{anysize}
\usepackage{appendix}
\usepackage{bm}
\usepackage{color}
\usepackage{dcolumn}
\usepackage{epsf}
\usepackage{epstopdf}
\usepackage{float}
\usepackage{graphicx}
\usepackage{mathrsfs}
\usepackage{mathtools}
\usepackage{multirow}
\usepackage{pstricks,pst-node,pst-text,pst-3d}
\usepackage{subfigure}
\usepackage{setspace}
\usepackage[all]{xy}

\DeclarePairedDelimiterX\braket[2]{\langle}{\rangle}{#1 \delimsize\vert #2}

\makeatletter
\newcommand\figcaption{\def\@captype{figure}\caption}
\newcommand\tabcaption{\def\@captype{table}\caption}
\makeatother

\newcommand{\e}{\epsilon}

\newcommand{\tr}{{\rm Tr}}

\begin{document}


\title{Robust Control of Single-Qubit Gates at the Quantum Speed Limit}

\author{Xi Cao}
\affiliation{Department of Automation, Tsinghua University, Beijing, 100084, China}
\author{Jiangyu Cui}
\affiliation{Central Research Institute, 2012 Labs, Huawei Technologies}
\author{Man Hong Yung}
\affiliation{Central Research Institute, 2012 Labs, Huawei Technologies}
\author{Re-Bing Wu}\email{rbwu@tsinghua.edu.cn}
\affiliation{Department of Automation, Tsinghua University, Beijing, 100084, China}


\begin{abstract}
	Fastness and robustness are both critical in the implementation of high-fidelity gates for quantum computation, but in practice a trade-off has to be made between them. In this paper, we investigate the underlying robust time-optimal control problem so as to make the best balance. Based on the Taylor expansion of the system's unitary propagator, we formulate the design problem as the optimal control of an augmented finite-dimensional system at its quantum speed limit (QSL), where the robustness is graded by the degree of series truncation. The gradient-descent algorithm is then introduced to sequentially seek QSLs corresponding to different orders of robustness. Numerical simulations for single-qubit systems show that the obtained time-optimal control pulses can effectively suppress gate errors (to the prescribed robustness order) caused by qubit frequency and field amplitude uncertainties. These results provide a practical guide for selecting pulse lengths in the pulse-level compilation of quantum circuits.
\end{abstract}

\maketitle

\section{Introduction}
{\it Precision} is the primary goal of control for gate implementation in quantum computation~\cite{Wang2011,Barends2014,Ballance2016}. On top of it, the designed control protocols need to be as {\it fast} as possible so that more gates can be performed within the system's finite coherence time. The control also needs to be {\it robust} against errors caused by noises and uncertainties in the system, so that high performance can be maintained with less frequent recalibration. The two important targets have stimulated extensive studies for robust quantum control~\cite{schlosshauer2019,arute2019,pellizzari1995} and time-optimal control~\cite{ribeiro2019accelerated,zhou2017accelerated}, respectively. However, combining them in a single design task, i.e., seeking controls that are not only fast but also robust, has been rarely considered.

\begin{figure}
	\begin{center}
		\includegraphics[width=0.9\columnwidth]{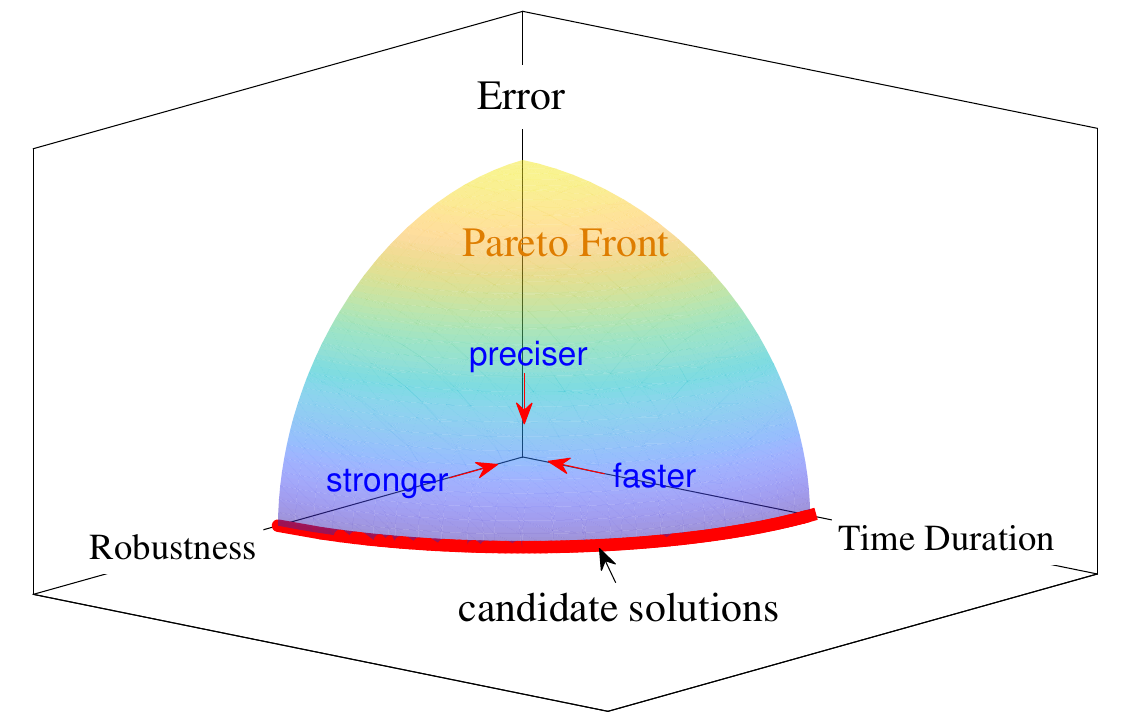}
	\end{center}
	\caption{Schematics of the Pareto front of the tri-objective optimization between precision, fastness and robustness. Since precision must be guaranteed in practice, we are concerned with the bottom edge that makes the best balance between robustness and fastness. The corresponding time duration of each point indicates the quantum speed limits with different degrees of robustness.  
	}\label{fig:pareto}
\end{figure}

The design task considering both fastness and robustness can be formulated as a tri-objective optimization problem acrossing the precision, the time duration and the robustness of the control pulse. As is shown in Fig.~\ref{fig:pareto}, the best compromised solutions constitute the Pareto front on which one index cannot be further improved without sacrificing the other two. Since the precision must be guaranteed with highest priority, the trade-off is mainly between the robustness and the fastness at the bottom edge of the Pareto front. For each specified degree of robustness, the corresponding time-duration at the edge corresponds to the minimum time for achieving high-precision robust gates, which is also called the quantum speed limit (QSL). 

In principle, the robust control of a quantum system with uncertainties is equivalent to the control of uncountably many determistic systems using a uniform control field, which can be proved to exist under certain Lie algebraic controllability conditions~\cite{Li2006}. To find such control fields, one can Taylor expand the ensemble control system into a series of interacting deterministic subsystems~\cite{VanDamme2017,Li2022}, which can be truncated to some order that speficies the desired robustness. This provides a framework for designing the fastest robust quantum control pulses for implementing dynamically corrected gates (DCG)~\cite{Khodjasteh2009}. For single-qubit systems, geometric restrictions on the truncated system can be applied to derive the analytic form of robust time-optimal control pulses~\cite{Zeng2018}. Alternatively, low-order robust time-optimal control solutions may be solved by Pontryagin Minimum Principle~\cite{VanDamme2017}. 

These known results focused on analytically solvable cases in which the system is controlled by a single field and is disturbed by a single type of uncertainty (e.g., field amplitude or qubit frequency uncertainties that causes bit-flip or phase-flip errors, respectively). Real quantum systems often involve multiple uncertainties that jointly affect the dynamics, under which circumstance multiple control fields (e.g., the microwave driving of superconducting qubits with both I/Q components)~\cite{krantz2019quantum,blais2021circuit} have to be applied to enhance the error suppression capability. Generally, no analytical solutions exist for such complicated cases and, to our knowledge, neither numerical studies can be found in the literature. 

In this paper, we will propose an algorithm for identifying robust QSLs and the corresponding robust time-optimal controls in a single-qubit system with two controls and two uncertainties. The obtained values of robust QSLs provide a useful guide for the choice of pulse length in consideration of uncertainties. The remainder of this paper is organized as follows. Section \ref{Sec:Model} presents the expansion-based model for robust time-optimal control problems, which grades the degree of robustness by the order of series truncation. Then, Sec.~\ref{sec:alg} analyzes the general properties of the corresponding QSL and the time-optimal controls, based on which a numerical algorithm is presented for searching QSLs and the corresponding robust control pulses. In section \ref{Simulation}, the proposed numerical algorithm is tested by a single-qubit system with two orthogonal controls and two uncertainty parameters. Finally, conclusion is made in Sec.~\ref{sec:conclusion}.

\section{The dynamical model of single-qubit control systems with uncertainties}\label{Sec:Model}
Let us start from an ideal single-qubit system that is resonantly driven by two orthogonal control fields. In the rotating frame, the evolution of the unitary propagator $U(t)$ obeys the following Schr\"{o}dinger equation:
\begin{equation}\label{eq:Model}
	\dot{U}(t)=-i\left[u_x(t)\sigma_x+u_y(t)\sigma_y\right]U(t)
\end{equation}
over some finite time interval $[0,T]$, where $U(t)$ is the unitary propagator of the qubit system and the Pauli matrices are
\begin{eqnarray*}
	\sigma_x & = & \frac{1}{2}\left(\begin{matrix}
		0 & 1\\
		1 & 0
	\end{matrix}\right), \\
	\sigma_y & = & \frac{1}{2}\left(\begin{matrix}
		0 & i\\
		-i & 0
	\end{matrix}\right), \\
	\sigma_z & = & \frac{1}{2}\left(\begin{matrix}
		1 & 0\\
		0 & -1
	\end{matrix}\right).
\end{eqnarray*} 
The time-dependent functions $u_x(t)$ and $u_y(t)$ are the in-phase and quadrature components of the control field subject to the following power constraint:
\begin{equation}
	u_x^2(t)+u_y^2(t)\leq {\Omega}^2,
\end{equation}
where the amplitude bound ${\Omega}$ is in unit of angular frequency. 

In this paper, we consider two typical types of uncertainties described as follows:
\begin{equation}\label{eq:ModelE}
\begin{aligned}
	\dot{U}(t;\epsilon_1,\epsilon_2)=&-i\large\{\epsilon_1 \sigma_z + (1+\epsilon_2)\left[u_x(t)\sigma_x+\right.
	\\&  \left.u_y(t)\sigma_y\right]\large\}U(t;\epsilon_1,\epsilon_2), 
\end{aligned}
\end{equation}
where the uncertainty paremeters $\epsilon_1$ (in unit of angular frequency) and $\e_2$ (dimensionless) characterize the drift of qubit frequency and the power fluctuation of control field, respectively. Throughtout this paper, we assume that the two parameters are slowly varying and hence can be treated as unknown constants during the course of evolution. 

The system model \eqref{eq:ModelE} is typical in the implementation of single-qubit gates. The operators $\sigma_z$ and $\sigma_x$ associated with the uncertainty parameters $\e_1$ and $\e_2$ can be taken as the infinitesimal generator of phase-flip and bit-flip errors that are broadly studied in stabilizer quantum error correction codes~\cite{google2021exponential}. The purpose of robust control is to reduce the error rates to be below the threshold for feedback-based error correction.  

To facilitate the following analysis, we divide \eqref{eq:ModelE} by a scaling factor $\Omega_0$ (in unit of angular frequency) on both sides, which leads to the following nondimensionalized system:
\begin{equation}
	\begin{aligned}
	\dot{U}(\bar t;\bar{\e}_1,\e_2) =&-i\large\{ \bar{\e}_1\sigma_z+(1+\e_2)\large[\bar{u}_x(\bar t)\sigma_x+\\ & \bar{u}_y(\bar t)\sigma_y\large]\large\}U(\bar t;\bar{\e}_1,\e_2), 
	\label{eq:model}
	\end{aligned}
\end{equation}
in which the rescaled time variable $\bar t=\Omega_0 t$ and uncertainty parameter $\bar\e_1=\e_1/\Omega_0$ are both dimensionless. The rescaled control functions
\begin{equation}\label{eq:transformation}
	\bar{u}_{x,y}(\bar t)=\Omega_0^{-1}u_{x,y}(\Omega t)	
\end{equation}
are also nondimensionalized with a dimensionless bound $\bar \Omega=\Omega/\Omega_0$. 

The above transformation shows that the robust time-optimal control problems can be analyzed and optimized with respect to a fixed bound $\bar \Omega$, and the obtained results can be extended by rescaling to cases with an arbitrary value of $\Omega$. Without loss of clarity, we will always assume the system is dimensionless and remove all the bars in the following discussions. 

\section{Searching algorithm for QSLs and robust time-optimal control pulses}\label{sec:alg}
In this section, we will analyze the properties of robust time-optimal controls and define the corresponding quantum speed limits, based on which a numerical algorithm will be proposed.

\subsection{The robust quantum speed limits}
The goal of robust time-optimal control in this paper is to find the shortest control pulse $u_{x,y}(t)$ so that $U(T;\e_1,\e_2)\equiv U_f$ for arbitrary $\e_1$ and $\e_2$. Since ${\e}_1$ and $\e_2$ are continuous parameters, this actually requires that the control must be able to steer {\it uncountably} many {\it non-interacting} subsystems (associated to different values of $\e_1$ and $\e_2$) to the same target $U_f$ at the same time $T$. 

Since the uncertainty parameters are small in most systems, we can Taylor expand $U(t;\e_1,\e_2)$ as the following series
\begin{equation}
U(t;\e_1,\e_2)=\sum_{k_1,k_2\geq 0}{\e}_1^{k_1}\e_2^{k_2} U_{k_1,k_2}(t),
\label{eq:expansion}
\end{equation}
where $\e_1$ has been nondimensionalized. In the small parameter regime, we only need to keep the first few dominant error terms. Correspondingly, the ensemble of uncountably many systems may be approximated by a finite dimensional system corresponding to the reserved terms. The order of series truncation naturally grades the robustness to be achieved. 

In this regard, a control $u_{x,y}(t)$ is said to be $(n_1,n_2)$-th order robust if it steers $U_{00}(T)$ to its target $U_f$ and meanwhile diminishes $U_{k_1,k_2}(T)$ for all other $0\leq k_1\leq n_1$ and $0\leq k_2\leq n_2$. Among all $(n_1,n_2)$-th order robust control pulses, the duration time $T_{n_1,n_2}$ of the shortest one is defined as the $(n_1,n_2)$-th order robust QSL. According to the transformation \eqref{eq:transformation}, the robust QSLs are proportional to the inverse of the control bound $\Omega$, which means that higher control bounds lead to shorter QSLs.

\subsection{Numerical algorithm}

The time-optimal control solutions are solvable for the lowest-order $n_1=n_2=0$, which are in form of either sinusoidal or constant functions~\cite{Albertini2015}. For general higher-order cases, we have to resort to numerical optimization as analytical solutions are unavailable. In this paper, we will use the following objective function for the search of robust time-optimal controls:
\begin{equation}\label{eq:cost}
	\begin{aligned}
	J[\{u_k(t)\}]=&F[U_{00}(T)] +{\rm tr}[U^\dag_{0,1}(T)U_{0,1}(T)]\\& +\cdots+{\rm tr}[U^\dag_{n_1n_2}(T)U_{n_1n_2}(T)],
\end{aligned}
\end{equation}
which includes the gate error 
\begin{equation}\label{eq:fidelity}
	F[U]=1-|{\rm tr}(U_f^\dag U)|^2/2^2
\end{equation}
and the higher-order error terms to be dimimished for $(n_1,n_2)$-th order robustness. 

The strategy for finding robust QSLs and their corresponding time-optimal controls is straightforward. Illuminated by our earlier works \cite{Tibbetts2012,Chen2015,Jacobs2016}, we start from a small time duration $T$ (shorter than the QSL) and optimize the control pulse using the GRAPE (GRadient-Ascent Pulse Engineering) algorithm~\cite{Khaneja2005}, after which we gradually increase $T$ and re-optimize the control pulse until $J$ is descreased to some error threshold $\epsilon$, and the corresponding critical time is recorded as the robust QSL $T_{n_1,n_2}$. Since the pulse shapes continuously vary with $T$, one can use the obtained optimal control pulse as the initial guess for the next round of optimization, which takes only a few iterations to update the control. 

To facilitate the gradient calculation, we first derive dynamical equations for $U_{k_1k_2}(t)$ by replacing \eqref{eq:expansion} into \eqref{eq:model}, which yields~\cite{VanDamme2017}:
\begin{equation}
	\dot{U}_{00}(t) =-i\left[ u_x(t)\sigma_x+u_y(t)\sigma_y\right]U_{00}(t), \\
\end{equation}
for the $(0,0)$-th order term, and 
\begin{eqnarray*}
	\dot{U}_{k_1k_2}(t) &=&-i\left[ u_x(t)\sigma_x+u_y(t)\sigma_y\right]U_{k_1k_2}(t)\\
	&&-i\sigma_z U_{k_1-1,k_2}(t)\\
	&&-i\left[ u_x(t)\sigma_x+u_y(t)\sigma_y\right]U_{k_1,k_2-1}(t),
\end{eqnarray*}
for higher-order terms. They can be further grouped into a compact form:
\begin{widetext}
	\begin{equation}\label{eq:Model_unified}
		\dot{\vec{U}}_{n_1n_2}(t)=-i\large\{\mathbb{L}_{n_1}\otimes \mathbb{I}_{n_2+1}\otimes \sigma_z+ \mathbb{I}_{n_1+1}\otimes(\mathbb{I}_{n_2+1}+\mathbb{L}_{n_2})\otimes\left[u_x(t)\sigma_x+u_y(t)\sigma_y\right]\large\}\vec{U}_{n_1n_2}(t),
	\end{equation}
\end{widetext}
where $\vec{U}_{n_1,n_2}(t)=[U_{00}(t),U_{01}(t),\cdots,U_{n_1n_2}(t)]^\top$ and 
\begin{equation}\label{eq:Model1}
	\mathbb{L}_n=\left(
	\begin{array}{cccc}
		0 & 0 & \cdots & 0 \\
		1 & 0 & \cdots & 0 \\
		\vdots & \vdots & \ddots & \vdots \\
		0 & \cdots & 1 & 0 \\
	\end{array}
	\right)\in\mathbb{R}^{(n+1)\times(n+1)}.
\end{equation}

The above model shows that the gradient calculation will be very expensive for high-order robust time-optimal controls due to the rapid increase of the systen dimension $N=2(n_1+1)(n_2+1)$. Regarding this, we can initiate the optimization from a low order for which the QSL is easy to find, after which we gradually increase the order and repeat the same procedure. In this way, robust QSLs and the corresponding time-optimal controls can be sequentially discovered from low to high orders.

The optimization efficiency can be further improved in the following ways. First, according to the Pontrygin Minimum Principle, the robust time-optimal controls satisfy $u_x^2(t)+u_y^2(t)={\Omega}^2$ (see appendix for details), i.e., the fastest control must be also the strongest. This property narrows down the search within the space of control fields in the following form
\begin{equation}\label{eq:NecessaryCondition}
	u_x(t)={\Omega}\cos\phi(t),\quad u_y(t)={\Omega}\sin\phi(t),
\end{equation}
where the phase function $\phi(t)$ is the free function to be optimized. 

Second, we derive a formulation (see Appendix \ref{sec:gradient}) to alleviate the computation cost of matrix exponentials for evaluating the evolution operator of \eqref{eq:Model_unified}, which rapidly rises with the robustness order. The formulation exploits the properties of Pauli matrices that enables a 4-10 times wall-time speed-up compared with the traditional Pade-approximation based method. The method also provides a simple formulation for precisely evaluating the gradient vector, while the formulation adopted in the standard GRAPE algorithm~\cite{Khaneja2005} is only a first-order approximation. The proposed accurate gradient-vector formulation guarantees the stability of the iterative optimization without increasing the computational cost.

\section{Simulation Results}
\label{Simulation}
In this section, we apply the proposed numerical algorithm to the search of robust time-optimal control pulses for single-qubit gates. For generality, we test the following four different one-qubit gates, 
\begin{eqnarray*}
	X & = & \left(\begin{matrix}
		0 & 1\\
		1 & 0
	\end{matrix}\right), \\
	Z & = & \left(\begin{matrix}
		1 & 0\\
		0 & -1
	\end{matrix}\right), \\
	S & = & \left(\begin{matrix}
		1 & 0\\
		0 & i
	\end{matrix}\right), \\
	H & = & \frac{1}{\sqrt{2}}\left(\begin{matrix}
		1 & 1\\
		1 & -1
	\end{matrix}\right). \\
\end{eqnarray*} 

The simulation is based on the nondimensionalized model \eqref{eq:model} with $\bar{\Omega}=\pi$, under which the time duration of a square $\pi$-pulse is unit. Let $\bar{T}_{n_1n_2}$ be the corresponding robust QSL, then the QSL for arbitray control bound $\Omega$ can be converted as ${T}_{n_1n_2}=\pi\Omega^{-1}\bar{T}_{n_1n_2}$. For each gate, we optimize robust time-optimal controls under three circumstances: (1) only the frequency is uncertain (corresponding to $n_1\neq 0$ and $n_2=0$); (2) only the field amplitude is uncertain (corresponding to $n_1=0$ and $n_2\neq 0$); and (3) both the frequency and the amplitude are uncertain (corresponding to $n_1\neq 0$ and $n_2\neq 0$). 

\begin{figure}
	\begin{center}
		\includegraphics[width=1\columnwidth]{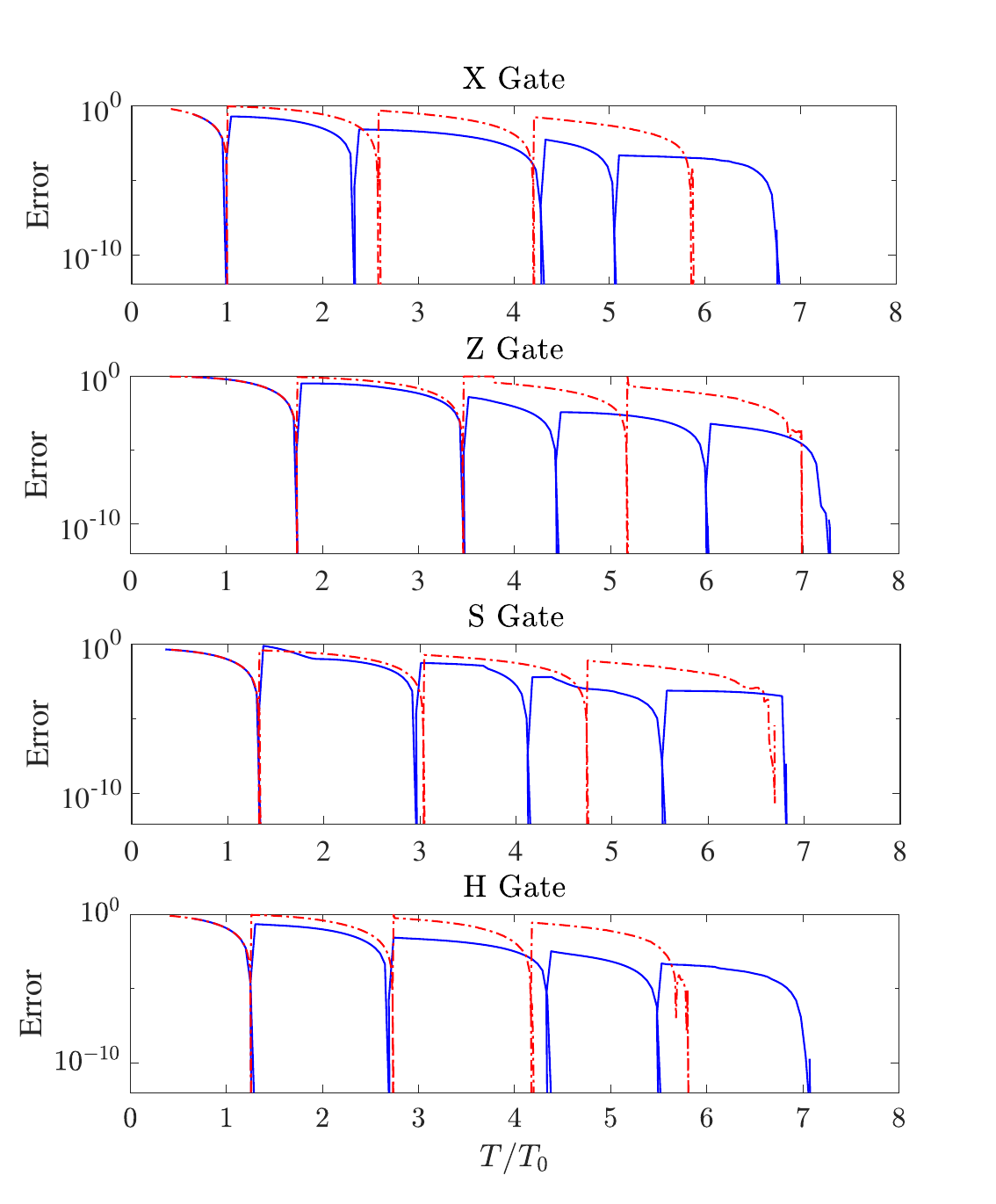}
	\end{center}
	\caption{The phase transitions during the optimization of robust time-optimal controls against the frequency (blue solid curves) and amplitude (red dash-dotted curves) uncertainties for gates $X$, $Z$, $S$ and $H$, respectively. The sharp dips from left to the right indicate the QSLs from zeroth-order to higher-order robust controls, where $T_0$ is the duration time of a rectangular $\pi$-pulse.
	}\label{fig:transition}
\end{figure}

\begin{figure}
	\begin{center}
		\includegraphics[width=1\columnwidth]{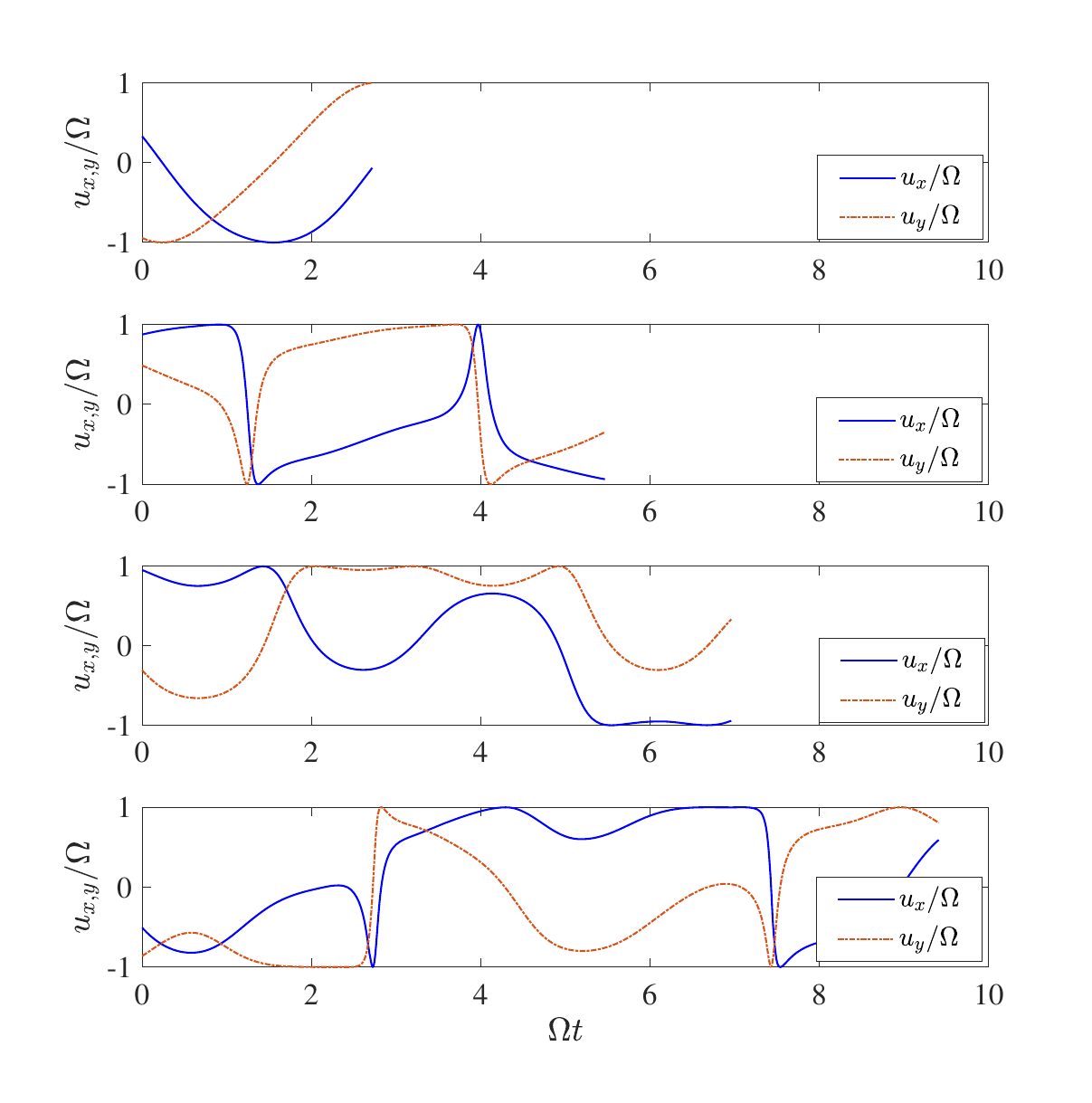}
	\end{center}
	\caption{The robust time-optimal control fields for the $Z$ gate with zeroth-oder to 3rd-order robustness, respectively, whose pulse durations increase with the robustness order.
	}\label{fig:field}
\end{figure}

We start the numerical optimization from a short duration time $T=0.3$ and from the lowest order $n_1=n_2=0$. Then, we gradually increase $T$ by $\Delta T=0.005$ in each round and update the field by minimizing the cost function \eqref{eq:cost} until the error threshold $\epsilon=10^{-10}$ is reached. The hitting time is then recorded as the corresponding QSL, after which we increase the robustness order by one and repeat the same procedure until the next QSL is detected. Figure~\ref{fig:transition} displays the transition of optimal $J$ value when increasing $T$ and the robustness order, where the sharp dips from left to the right indicate the robust QSLs from zeroth-order to higher orders. The resulting robust time-optimal control fields for $Z$ gate are displayed in Fig.~\ref{fig:field}. 

During the optimization, the pulse duration time is always below the QSL and thus the control system \eqref{eq:model} remains uncontrollable. The lack of controllability may cause unwanted local suboptima that traps the search away from globally optimal solutions~\cite{Rabitz2004,Rabitz2005}. This did happen in our simulations, which is severer when the robustness order is high. Under such circumstances, we need to run the algorithm from different initial guesses and select the shortest transition time as the approximated robust QSL.

	\begin{table*}[t]
		\centering
		\caption{The robust QSLs for single-qubit gates ($X$, $Z$, $S$ and $H$) against frequency or/and amplitude uncertainties.}
		\label{table1}
		\begin{spacing}{1.5}
			\begin{tabular}{|c||c||c|c|c|c||c|c|c||c|c|}
		\hline
	\multirow{2}{*}{Gate}&\multirow{2}{*}{~~$\bar T_{00}$~~} & \multicolumn{4}{c||}{Frequency}&\multicolumn{3}{c||}{Amplitude}&\multicolumn{2}{c|}{Freq.+Amp.}\\
		\cline{3-11}		
	\multirow{2}{*}{} &\multirow{2}{*}{}    & ~~$\bar T_{10}$~~ & ~~$\bar T_{20}$~~& ~~$\bar T_{30}$~~& ~~$\bar T_{40}$~~& ~~$\bar T_{01}$~~& ~~$\bar T_{02}$~~& ~~$\bar T_{03}$~~& ~~$\bar T_{11}$~~& ~~$\bar T_{22}$~~\\
		\hline		
		$X$	& 1.00	& 2.33 & 4.28   & 5.04 & 6.72  & 2.58 	& 4.21 	& 5.85  &  4.44 &  8.22  \\ 
		\hline		
		$Z$	& 1.74	& 3.48  & 4.43 & 5.99  & 7.19  & 3.46   & 5.17	& 6.91  &  5.34   &  8.78 \\
		\hline		
		$S$	& 1.32	& 2.97  & 4.12 & 5.53  & 6.71  & 3.04	& 4.74	& 6.48  &  4.83   &  8.11 \\ 
	\hline		
		$H$	& 1.25	& 2.69  & 4.34 & 5.47  & 7.00  & 2.73	& 4.18	& 5.81  &  4.89   &  8.83 \\ 
	\hline
		\end{tabular}
		\label{table_MAP}
	\end{spacing}		
\end{table*}

\begin{figure}
	\begin{center}
		\includegraphics[width=1\columnwidth]{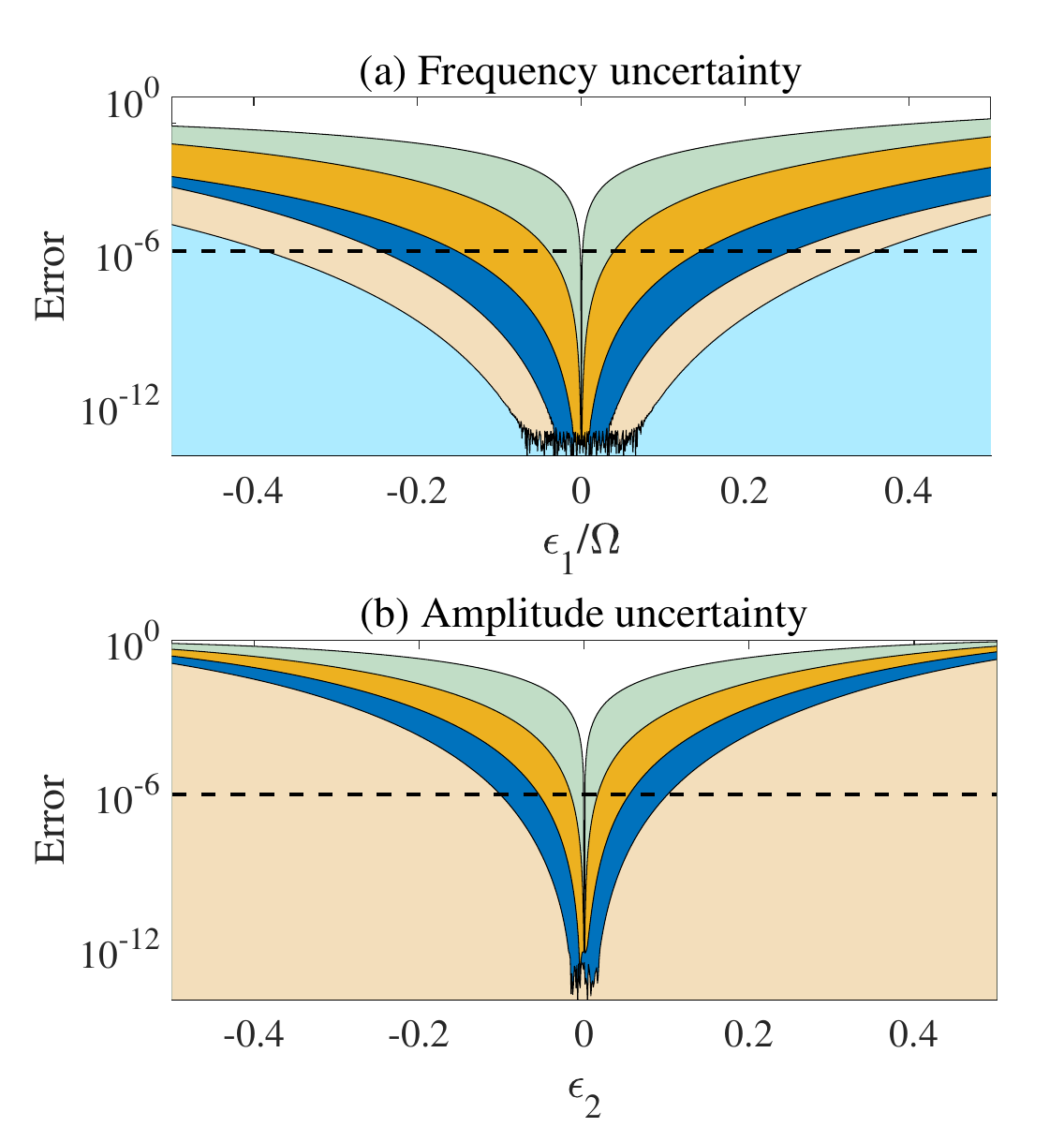}
	\end{center}
	\caption{ The gate error versus uncertainty parameter under robust time-optimal controls for $Z$ gate: (a) frequency uncertainty (from top to bottom: zeroth-order to fourth-order); (b) amplitude uncertainty (from top to bottom: zeroth-order to third-order). 
	}\label{fig:error}
\end{figure}

Table~\ref{table_MAP} enlists the obtained QSLs for relatively low orders, whose corresponding controls can meet the demand of most applications. The calculation of higher-order QSLs is much harder and also unnecessary because they may be longer than the qubit's coherence time. To test the robustness performance of the obtained time-optimal controls, we evaluate the gate error~\eqref{eq:fidelity} using the uncertain model \eqref{eq:ModelE} when the frequency and amplitude uncertainty parameters are evenly sampled within $-0.5\leq \epsilon_1 \leq 0.5$ and $-0.5\leq \epsilon_2 \leq 0.5$. Taking the Z-gate for example, the dependence of the gate error on the uncertainty parameter is depicted in Fig.~\ref{fig:error} for different orders of robustness. It can be seen that higher-order robust control can maintain high precision in a wider range. For examples, the 3rd-order robust time-optimal controls can suppress the error down below $10^{-6}$ (a typical threshold for quantum error correction) within the regime $|\e_1|\leq 0.26$ for frequency uncertainty and $|\e_2|\leq 0.10$ for amplitude uncertainty. This is to say that, to maintain the $10^{-6}$ precision under 10MHz bound on the driving field, the frequency offset is allowed to drift by up to 2.6MHz abd the field amplitude is allowed to shift by up to $\pm 10\%$.

\begin{figure*}
	\begin{center}
		\includegraphics[width=1\columnwidth]{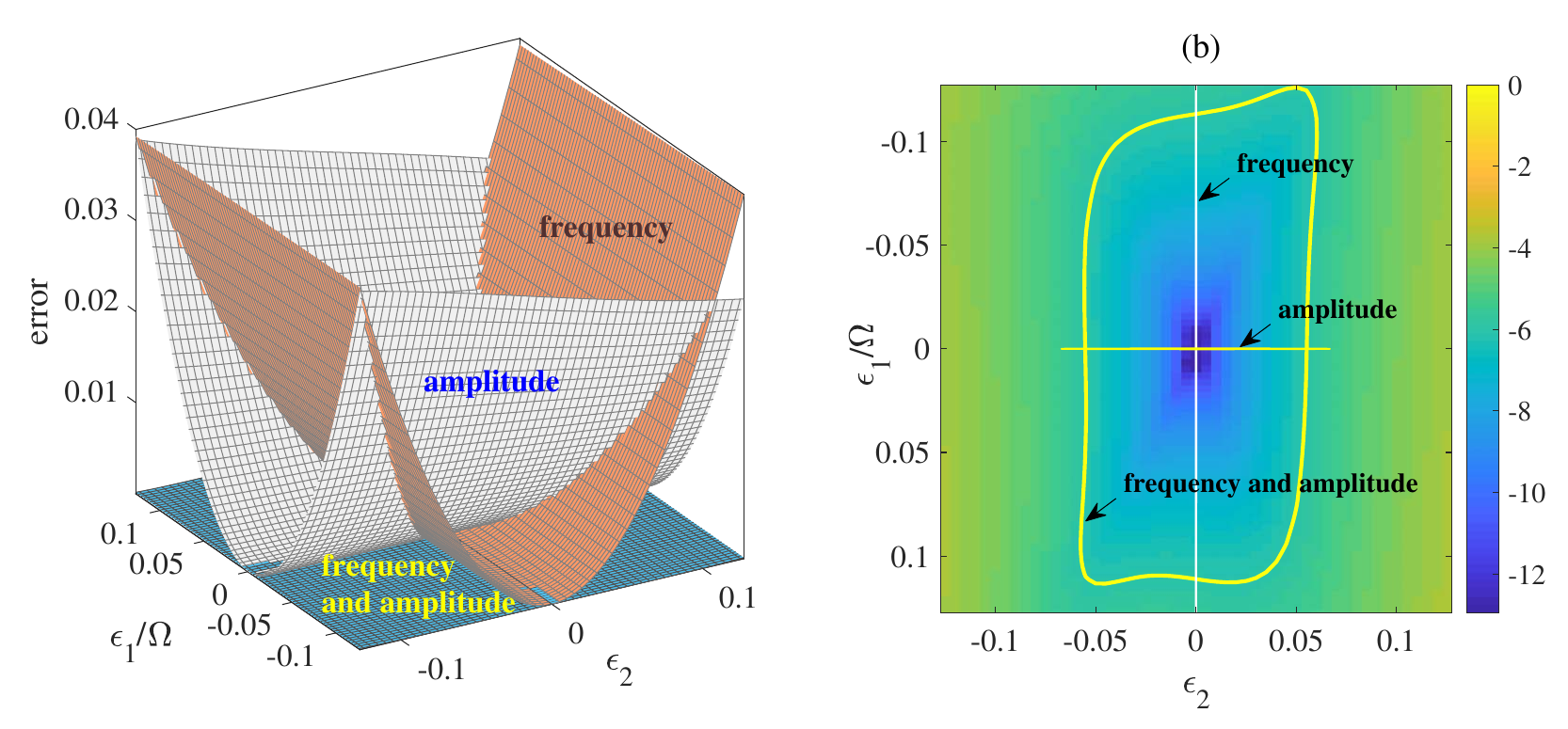}
	\end{center}
	\caption{The gate error versus frequency or/and amplitude uncertainties under $(2,2)$-th, $(2,0)$-th and $(0,2)$-th order robust time-optimal controls: (a) 3D plot; (b) the color map of the logrithm of the gate error and the contour curves at the level set $10^{-6}$.
	}\label{fig:error1+2}
\end{figure*}

We also test the performance of $(2,2)$-th order robust time-optimal control that resists simultaneous frequency and amplitude uncertainties. The dependence of the gate error on the two parameters is visualized by the 3D plot in Fig.~\ref{fig:error1+2}(a). The flat landscape indicates that the control can dynamically correct errors simulataneously induced by the two uncertainties. By contrast, the 2nd-order robust time-optimal controls with respect to individual frequency (or amplitude) uncertainty is extremely fragile to the amplitude (or frequency) uncertainty. To see this more clearly, we draw in Fig.~\ref{fig:error1+2}(b) the contour plots at the level-set of $10^{-6}$. The area enclosed by the contour curves indicate shows the overwhelmingly superior robustness of the $(2,2)$-th order control comparing to those of single-parameter second-order robust controls. However, it should be noted that the simultaneous robustness against two uncertainties is achieved at the price of a longer pulse duration $\bar{T}_{22}=8.78$ in comparison with $\bar{T}_{20}=4.43$ and $\bar{T}_{02}=5.17$.

\section{Concluding Remarks}\label{sec:conclusion}
To conclude, we have presented the concept of robust QSL and proposed a numerical algorithm for designing robust time-optimal control pulses that balance between control speed and robustness. Numerical simulations are performed for single-qubit quantum gates against frequency and field amplitude uncertainties, which sequentially discover QSLs from low to high orders and their corresponding robust time-optimal controls. Their excellent robustness performance are verified by numerical tests. To our knowledge, this is the first time that robust time-optimal quantum controls are designed for systems with multiple control fields and multiple uncertainty parameters. 

The QSLs enlisted in Table \ref{table_MAP} provide a useful guidance for selecting pulse durations in the pulse-level compilation of quantum circuits, so that the entire time-evolution can be minimized while maintaining error robustness. This can effectivly reduce the error-per-gate (EPG) rate, which is pivotal for the implementation of error-correctible quantum computing algorithms. 

The constructed model can be directly extended to more general systems, as well as the proposed optimization algorithm. In generic scenarios, it is important to investigate the system's controllability so as to understand to what extent one can correct the errors in complex uncertain quantum systems. For numerical optimization, the computational overhead will be a big challenge. We believe that there is much room left for developing more efficient and more stable algorithms. For example, one can alter the expansion of $U(t;\e_1,\e_2)$ under other polynomial basis of $\e$ instead of the monomial basis adopted in Eq.~\ref{eq:expansion}~\cite{Li2022}, which may lead to quicker convergence depending on the characteristics of the parameter distribution. Also, the methodology of quantum brachistchrone can also be applied for our purposes. These possibilities will be further explored in our future studies.

\acknowledgements
This work is supported by NSFC (Grants No.~61833010 and No.~62173201) and a grant from HUAWEI.

\appendix
\section{Proof of Property 1}\label{sec:poytry}
This property is based on the Pontryagin Minimum Principle (PMP) in optimal control theory. For convenience, we rewrite the unified model \eqref{eq:Model_unified} as 
\begin{equation}\label{eq:ModelPMP1}
	\dot{\vec{U}}_{n_1n_2}(t)=-i\left[H_0 +u_x(t)H_1+u_y(t)H_2\right]\vec{U}_{n_1n_2}(t),
\end{equation}
where 
\begin{eqnarray*}
	H_0 &=& \mathbb{L}_{n_1}\otimes \mathbb{I}_{n_2+1}\otimes \sigma_z,  \\
	H_1 &=& \mathbb{I}_{n_1+1}\otimes(\mathbb{I}_{n_2+1}+\mathbb{L}_{n_2})\otimes \sigma_x,  \\
	H_2 &=& \mathbb{I}_{n_1+1}\otimes(\mathbb{I}_{n_2+1}+\mathbb{L}_{n_2})\otimes \sigma_y. 
\end{eqnarray*}
PMP claims that time optimal controls must minimize the following pseudo Hamiltonian
\begin{equation}\label{}
	\mathcal{H}=1+ Re\tr\left\{\vec{V}_n^\dag(t)\left[H_0 +u_x(t)H_1+u_y(t)H_2\right]\vec{U}_n(t)\right\},
\end{equation}
where the adjoint-state $\vec{V}_n(t)$ satisfies the following differential equation:
\begin{equation}\label{eq:ModelPMP2}
	\dot{\vec{V}}_n(t)=-i\left[H^\dag_0 +u_x(t)H^\dag_1+u_y(t)H^\dag_2\right]\vec{V}_n(t),
\end{equation}
along the optimal trajectory of $\vec \theta(t)$ and $\vec V(t)$. This implies that 
\begin{equation}\label{eq:ModelPMP3}
	u_x(t)=-{\Omega}\cos\phi(t),\quad u_y(t)=-{\Omega}\sin\phi(t),
\end{equation}
where 
\begin{equation}
	\phi(t)=\arctan\frac{Re\tr\left[\vec{V}_n^\dag(t)H_2\vec{U}_n(t)\right]}{Re\tr\left[\vec{V}_n^\dag(t)H_1\vec{U}_n(t)\right]}
\end{equation}
as long as 
\begin{equation}
\begin{aligned}
\left\{Re\tr\left[\vec{V}_n^\dag(t)H_1\vec{U}_n(t)\right]\right\}^2+&\\ \left\{ Re\tr\left[\vec{V}_n^\dag(t)H_2\vec{U}_n(t)\right]\right\}^2&\neq0.
\end{aligned}
\end{equation}
The above result indicates that the time-optimal control always run with full power exerted. 

Theorectically, it is also possible that $\nu_1^2(t)+\nu_2^2(t)=0$, under which circumstance the optimal control can be determined by minimizing the pseudo Hamiltonian as above. Such piece of control is called a singular arc. We are not able to prove that singular arcs do not exist, but from our numerical simulations it is always the case. 



\section{The evaluation of gradient}\label{sec:gradient}
To calculate the cost function and its gradient, we need to calculate the propagator as a time-ordered exponential of the time-dependent Hammiltonian:
\begin{equation}
	\begin{aligned}
V(T) = &\mathcal{T}\exp\left\{-i\int_0^T \large[H_0+{\Omega}\cos\phi(t)H_1+\right.\\&\left.{\Omega}\sin\phi(t)H_2\large]dt\right\}.
\end{aligned}
\end{equation}
When the control pulse is piecewise constant, we can always decompose the propagator as $V(t)=V_N\cdots V_2V_1$, where
\begin{equation}
	\begin{aligned}
	V_j =& \exp\large\{-i\large[K_1\otimes \sigma_z+K_2\otimes({\Omega}\cos\phi_j \sigma_x+\\&{\Omega}\sin\phi_j\sigma_y)\large]\Delta t \large\},
	\end{aligned}
\end{equation}
where $K_1=\mathbb{I}_{n_1+1}\otimes(\mathbb{I}_{n_2+1}+\mathbb{L}_{n_2})$ and $K_2=
\mathbb{I}_{n_1+1}\otimes(\mathbb{I}_{n_2+1}+\mathbb{L}_{n_2})$. With this decomposition, the gradient can be computed as  
\begin{equation}
	\frac{\partial V(T)}{\partial \phi_j}= V_N\cdots\frac{\partial V_j}{\partial \phi_j}\cdots V_1.
\end{equation}

In most numerical tools, Pade approximation is broadly applied to calculte the matrix exponentials, and this is the most expensive part of the gradient-based algorithm. In the following, we show that, by exploiting the speical structure of the single-qubit control system, the matrix exponential can be computed more efficiently. This is based on the following fact that
\begin{equation}
	\begin{aligned}
\left[K_1\otimes \sigma_z+K_2\otimes({\Omega}\cos\phi_j \sigma_x+{\Omega}\sin\phi_j\sigma_y)\right]^2 \\= (K_1^2+{\Omega}^2K_2^2)\otimes \mathbb{I}_2
\end{aligned}
\end{equation} 
is independent of $\phi$. Apply this property to the Taylor expansion of the matrix exponential, we have
\begin{eqnarray*}
	V_j &=& \exp\large\{-i\large[K_1\otimes \sigma_z+K_2\otimes({\Omega}\cos\phi_j \sigma_x+\\&&{\Omega}\sin\phi_j\sigma_y)\large]\Delta t \large\}\\
	&=&  C(\Delta t)\otimes \mathbb{I}_2-i \large[S(\Delta t)K_1\otimes \sigma_z+S(\Delta t)K_2\otimes\\&&({\Omega}\cos\phi_j \sigma_x+{\Omega}\sin\phi_j\sigma_y)\large]\Delta t.
\end{eqnarray*} 
where 
\begin{eqnarray}
C(\Delta t)&=&\sum_{m=0}^\infty\frac{(-\Delta t)^m}{(2m)!}(K_1^2+{\Omega}^2K_2^2)^m  ,\\
S(\Delta t)&=& \sum_{m=0}^\infty\frac{(-\Delta t)^m}{(2m+1)!}(K_1^2+{\Omega}^2K_2^2)^m.	
\end{eqnarray}
Because $C(\Delta t)$ and $S(\Delta t)$ is only dependent on $\Delta t$ and ${\Omega}$, they can be precalculated and stored as constant matrices during the optimiztion. This can greatly improve the numerical efficiency.

Since the propagator is only linearly dependent on $\cos\phi$ and $\sin\phi$, we can easily calculate the term 
\begin{equation}
\frac{\partial V_j}{\partial \phi_j}= -i S(\Delta t)K_2\otimes(-{\Omega}\sin\phi_j \sigma_x+{\Omega}\cos\phi_j\sigma_y)\Delta t
\end{equation}
gradient. Note that the gradient vector calculated in this way is without any approximation, while in standard GRAPE algorithm~\cite{Khaneja2005} the gradient evaluation 
\begin{equation}
	\frac{\partial V_j}{\partial \phi_j}\approx -i \left[K_2\otimes(-{\Omega}\sin\phi_j \sigma_x+{\Omega}\cos\phi_j\sigma_y)\right]V_j\Delta t
\end{equation}
is based on the first-order approximation.


	\bibliography{rTOC}

\end{document}